\documentclass{epl}

\newcommand{\beq}{\begin{equation}}
\newcommand{\eneq}{\end{equation}}
\newcommand{\met}{\frac{1}{2}}

\title{Possibility of Two-channel Spin $\frac{1}{2} $ Kondo
           Conductance  in a Quantum Dot}
\author{D. Giuliano \inst{1,2} \and B. Jouault \inst{1,3} \and 
A. Tagliacozzo \inst{1,2}}
\institute{
  \inst{1} Istituto Nazionale di Fisica della Materia (INFM),
         Unit\'a di Napoli \\
  \inst{2}  Dipartimento di Scienze Fisiche Universit\`a di Napoli
         "Federico II " \\ 
         Monte S.Angelo - via Cintia, I-80126 Napoli, Italy \\
  \inst{3} GES, UMR 5650, Universit\'e Montpellier $II$,
         34095 Montpellier Cedex 5, France
}

\begin{document}

\maketitle

\begin{abstract}
By combining exact diagonalization with scaling method,
we show that it is possible to realize two channel spin $\met $ Kondo
(2CK) conductance in a  quantum dot at Coulomb Blockade, with an odd number of
electrons and with contacts in a pillar configuration, as an applied
orthogonal magnetic field $B$ is tuned at an appropriate level crossing.
\end{abstract}

A Quantum Dot (QD) weakly coupled to the contacts and tuned   in a  valley
between two conduction peaks (Coulomb Blockade (CB)), is insulating if
its charging energy is larger than the thermal energy \cite{aleiner}.
However, when the number of electrons at the dot, $N$, is odd, below  a
characteristic temperature scale $T_K$, a  strongly
correlated state  between the dot  and the contacts  sets in, and the typical
Kondo resonance in the conduction electron spectral density  builds up at
the  chemical potential of the contacts  $\mu$.
The striking result is that the  linear conductance  increases in the CB
valley, when the temperature $T$ is lowered,  up to the unitarity limit
$2e^2 / h$  for $T=0$ \cite{goldhaber,silvano0}.
If $N$ is even, the ground state (GS) of the QD is usually a spin singlet
and the ordinary Kondo effect cannot occur. Nevertheless, it has been shown
that level crossing between states at different  $S$ induced by a magnetic
field $B$ orthogonal to the dot can restore the degeneracy required for
the Kondo effect to take place  \cite{silvano,noi,pg}.
The  occurrence of Kondo physics in quantum dots was predicted long ago in
analogy to  magnetic impurities in diluted metal alloys at very low
temperatures \cite{raik}. A dot at CB acts as a single magnetic impurity, but
under controlled  experimental conditions.

\begin{figure}
\onefigure[width=0.9\linewidth]{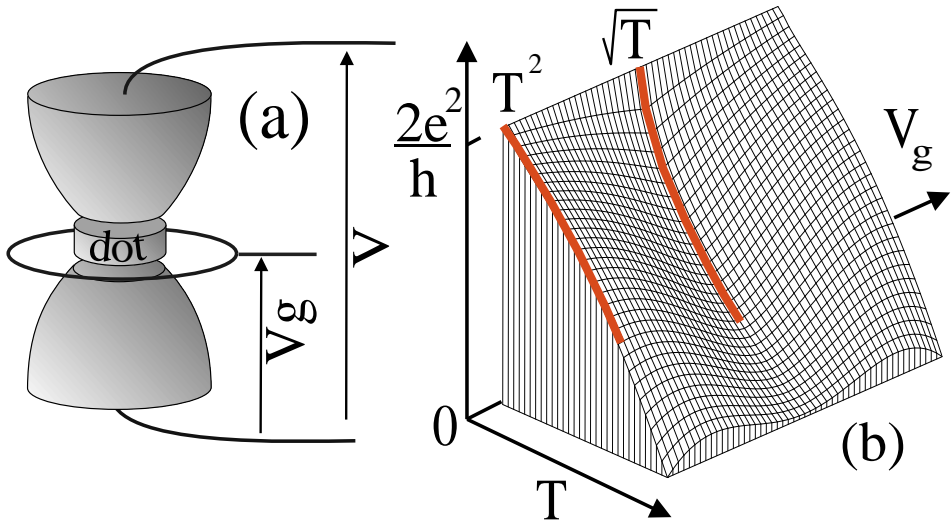}
\caption{(a): Sketch of the dot. (b)  Sketch of the conductance as a function
of the temperature $T$ and the gate voltage  $V_g$.
At 2CK non Fermi liquid point, the  conductance shows a $\sqrt{T}$ 
dependence. }
\label{uno}
\end{figure}

A realistic description of the Kondo effect  in alloys has to take into
account that the impurity  states carry also orbital angular momentum
together with spin momentum. Hence, electrons with different orbital
momentum can access the impurity and both multi-orbital 1-channel Kondo
Effect (1CK) and many-channel Kondo effect (MCK) can take
place. Total angular momentum is  conserved in the scattering,  as it is
appropriate for an atomic impurity in an isotropic metal \cite{blandin}.

Similarly to   the ordinary 1CK, MCK shows a logarithmic low-temperature
raise in the resistivity before the perturbative expansion breaks down
($T \sim T_K$). However, if the number of channels $k$  $> 2S$ (overscreening),
the physics of the two systems at $T \ll T_K$ is
dramatically different. The overscreened impurity   interacts again
antiferromagnetically with the other conduction  electrons. Accordingly, the
strongly coupled antiferromagnetic (AF) fixed point  becomes unstable at
$T=0$  and moves to intermediate coupling. The  Fermi-liquid phase
breaks down and leading corrections to the resistivity $\rho (T)$
are characterized by a non-Fermi liquid  square-root temperature dependence
\cite{affleck}. The prototype of this situation is an $S=\met$ impurity
interacting with two channels of conduction electrons.

 Possibly, 2CK zero-bias conductance anomaly has been detected in  transport
measurements in  $Ti-$ and $Cu-$ nanoconstrictions with controlled doping,
with $T_K$ between 0.5 $K$ and 4.6 $K$ \cite{ralph1}.  Also, interstitial
atoms in glassy metals could be described as tunneling two-level systems
(TLS) and they could  give rise to 2CK by scattering with conduction
electrons \cite{zawad}. However,  a careful estimate of  the Kondo
temperature in these materials  is still a widely debated question and
appears to give a value  below  $10^{-4}$ $K$\cite{zawa6}.

In this letter we propose to search for 2CK spin $\met$ in a  quantum dot
(QD) biased at Coulomb blockade with odd $N$ and total spin $S=1/2$. We
discuss the geometry, the working conditions, and  the  most appropriate energy
level structure to achieve 2CK  within  the limits of present technology.

We consider a QD in a pillar configuration with cylindrical symmetry
about the $z$-axis (see Fig.1 $a)$ ). We label
dot states with  total energy $^N\!E_i $  ( $i=0$ for the GS)
and  total angular momentum $M$ along the $z$ axis.
By means of single-electron tunneling with the contacts, the dot can exchange
orbital momentum $m$ and/or spin $\sigma$ 
with conduction electrons. We describe a particular
scenario derived  from results of an exact diagonalization of electrons in
a two-dimensional parabolic confining potential with Coulomb interaction
energy scale $U$ \cite{jouault}. The addition energies of the dot can be
shifted with respect to $\mu$ by tuning a  gate voltage  $V_g$. The
field $B$ orthogonal to the dot can induce strong orbital effects, which
may eventually give raise to crossings between states with different $M$
and $S$. The dot is tuned at CB with  $N=5$ at $B^*$ where  two
$S = \met $ levels with $M=4$ and $M=6$ cross (Fig.2 ).

\begin{figure}
\centering \includegraphics*[width=0.7\linewidth]{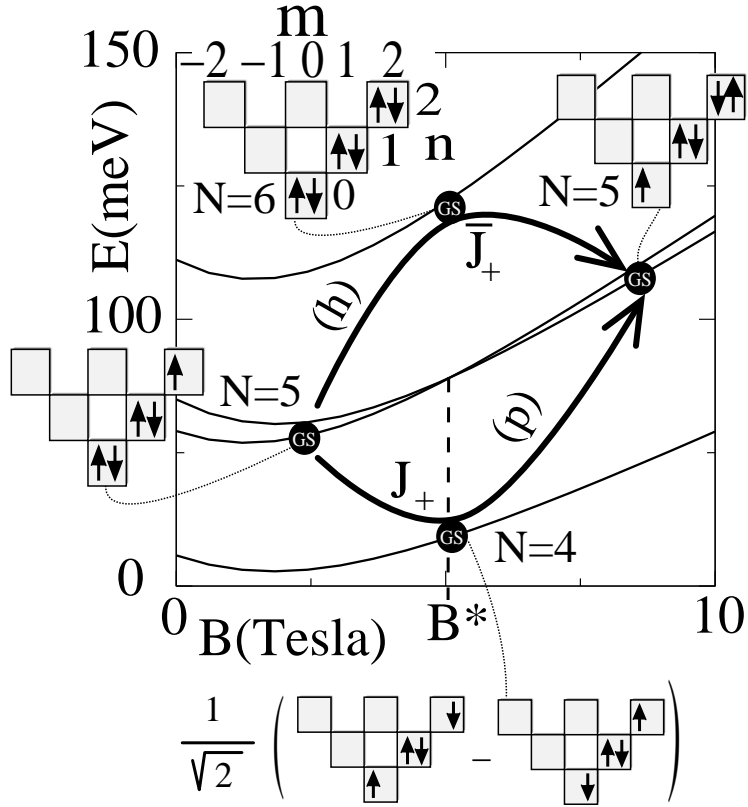}
\caption{  The lowest
dot energy levels  $^NE_i$, obtained by exact diagonalization, are
plotted as a function of the magnetic field $B$. At  the degeneracy point
 $B^ * \simeq$ 5 Tesla two $ S = \met $ levels for N=5 cross. The dot switches
between the two states via  electron cotunneling to and from the contacts
in the particle ($p$) and the hole ($h$) channel ( marked by thick arrows:
the  starting and
ending points are displaced from $B^*$ for clarity).  Intermediate
virtual many-body states  with N=4,6  are singlets. Symbols are explained
in the text.}
\label{due}
\end{figure}

Within  the subspace $\Xi $ spanned by the  states  $|N,M,S_z \rangle = $
$| 5, 4, \pm  \met  \rangle$  and  $| 5, 6,\pm \met  \rangle$,  the QD
may scatter a conduction electron with spin $\sigma$ from $m=0$ to $m=2$,
and vice versa. In this respect, the dot acts
as an impurity spin $S=3/2$  in  interaction with two channels of
conduction electrons (In our case the channel index is  the spin index
$\sigma = \uparrow , \downarrow$).
This would attempt to realize underscreening of the dot's spin
with Kondo temperature $T_{\rm cross}$. 
However, if Zeeman spin splitting (ZSS)  $\Delta_Z$ 
$>T_{\rm cross}$ ($k_B = \hbar =1$ throughout the paper),
two of the dot states are ruled out at temperatures $T$
such that  $\Delta_Z > T > T_{\rm cross}$.
The system may flow toward an overscreened $S=\met$ 2CK fixed point,
provided its Kondo temperature $T_{2CK}$ is such that $T_{2CK}
> T_{\rm cross} $. (The ZSS should  not  affect the density
of conduction electron states of both spin polarization at $\mu$.)

We now discuss for which values of the parameters the system may exhibit
2CK  and estimate  the corresponding Kondo temperature (see Fig.3) by
combining exact diagonalization results for an isolated dot with
$N=4,5,6$ electrons with scaling approach. We construct the even-parity
single-particle states labeled by $q m\sigma $  out of the wavefunctions
  $e^{i(k_{F,lm}+q)z}\varphi _{\epsilon ,q,m,\sigma}(\rho )$ \cite{parity},
delocalized in the $z$-direction,  with $\epsilon \sim \mu$ and
$q \ll k_{F,lm}$ ($ k_{F,lm}$ is the Fermi wavevector  of the 1-d subbands
of motion along $z$, labeled by $l$ and linearized about $\mu$). We denote the
corresponding operators by $b_{q m \sigma}$,  $b^\dagger_{q m \sigma}$
and drop the energy and subband index for simplicity. The Hamiltonian for
the leads is given by $H^b = \sum _{qm\sigma } (\epsilon _{qm} -\mu )
b^\dagger_{qm\sigma } b _{qm\sigma } $.

Dot states can be expanded in the basis of Slater determinants of single
particle orbitals ($n,m$ ) associated to the fermion operators
$d_{n m\sigma}, d^\dagger_{n m\sigma}$ (in a in-plane parabolic confining
potential,
$n=0,1,..\infty$ and $m= -n, -n+2,...,n-2,n$). In Fig.2 we represent the
determinants with $S_z =\met$ giving the largest contribution to the
$N=4,5,6$ GS, by filling a sequence of boxes. $n$ increases by one along the
vertical  rows of boxes , while $m$ increases along the  horizontal rows. A
large magnetic field $B \parallel \hat{z}$ favors the occupancy of positive
$m$ states. All the states have the lowest possible $S$, since
 Hund's rule does not apply
in the presence of a rather strong $B$.
The larger is  $U$, the higher in energy are $N=4,6$-triplet states.
Therefore, we neglect virtual transitions to triplets, since $B$
should be cranked up a lot in order to induce transitions to higher
spin states of the dot.

Within the subspace $\Xi$, the Hamiltonian of the isolated dot is:

\beq
H_d = K\sum _\sigma S^z_{\sigma \sigma} + \Delta _Z ( P_{\uparrow \uparrow}
 -  P_{\downarrow \downarrow} )
\label{hdot}
\eneq
\noindent
In Eq.(\ref{hdot}), we have introduced the following dot operators:

\[
 P_{\sigma \sigma^{'}} = \frac{1}{2} \sum_\alpha | 5 \alpha \sigma
\rangle \langle 5 \alpha \sigma^{'} | \; ; \;\;\; S^a_{\sigma \sigma^{'}} =
\frac{1}{2} \sum_{ \alpha \alpha^{'}} | 5 \alpha \sigma \rangle
\tau^a_{\alpha \alpha^{'}} \langle 5 \alpha^{'} \sigma^{'} |
\]
\noindent
$\alpha$ labels states with different  $M$ and $\tau ^a_{\alpha \alpha '}$
denote the Pauli matrices with $a=1,2,3$. The first  term  of
eq.(\ref{hdot}) is a possible detuning of $B$ from $B^*$, which breaks the
degeneracy of the dot states at different $M$  and is ignored in the following.
The second term is the ZSS term.

The model hamiltonian  for the  tunneling between dot and leads is
$H_t = \Gamma \sum _{nqm\sigma} d^\dagger _{nm\sigma}b_{qm\sigma} +h.c.$
To lowest order in $|\Gamma |^2 $,  $H_t$ connects dot states $\in \Xi$
to states with $N\pm 1$. At  CB, $ ^6\!E_0 -^5\!E_0 < \mu < ^5
\!E_0 -^4\!E_0$ and the two most relevant cotunneling processes  giving
raise to tunneling of  the dot  between states with $M=4$ and $M=6$ are
marked
by thick solid arrows in Fig.2. The labels $p,h$ refer to whether
the lead intermediate state has one extra electron ({\sl particle}), or
one extra hole ({\sl hole}).

Let us denote the states of the contacts by $| [ {\cal{N}} ],\phi )$, where
${\cal{N}}$ is the number of electrons  and  $\phi$ is a collective label for
all the appropriate quantum numbers. Generalized  magnetization  vector
operators acting on the contacts at the dot site $z=0$ in the perturbative
$h,p$-process \cite{nota} are given by:

\begin{eqnarray}
\begin{array}{c} \vec{m}_{\sigma ' \sigma }^{p} \\
 \vec{m}_{\sigma  \sigma ' }^{h} \end{array} \biggr\}
 = \frac{1}{2} \sum^{ q  m m'}_{ \phi \phi '\phi ''}
\biggl |  [{\cal{N}} \pm 1  \mp \{ q
m \sigma \}],\phi \biggr  )  \biggl ( [{\cal{N}} \pm 1], \phi ' \biggr |
\nonumber\\
\: \:\: \vec{\tau}_{m m^{'}}  \: \:\:
 \biggl | [{\cal{N}} \pm \{q m^{'}
\sigma^{'}\}] ,\phi ' \biggr ) \biggl ( [{\cal{N}}],\phi '' \biggr  | \:
\label{mag}
\end{eqnarray}
\noindent
Generalized scalar charge density operators, $ \rho_{\sigma \sigma^{'}}^{p,
h}$, are obtained by substituting $\vec{\tau}_{m m^{'}}$ with $\delta _{m
m^{'}}$ in Eq.(\ref{mag}). A  Schrieffer-Wolff transformation provides the
effective Hamiltonian restricted to the subspace $\Xi$:

\[
 H_{\rm Eff} = J \sum_\sigma [ P_{\sigma \sigma} \rho_{\sigma \sigma}^p +
\vec{S}_{\sigma \sigma}\cdot \vec{m}_{\sigma \sigma}^{p} ] +
\bar{J} \sum_\sigma [- P_{\sigma \sigma} \rho^h_{\bar{\sigma}
\bar{\sigma}}
\]

\beq
 + \vec{S}_{\sigma \sigma}\cdot \vec{m}^{h}_{\bar{\sigma}
\bar{\sigma}} ] +
Y \sum _{x=p,h} \sum_\sigma
\vec{S}_{\sigma \bar{\sigma}} \cdot \vec{m}_{ \bar{\sigma}\sigma }^{x} \;\;
;
\label{eff4}
\eneq
\noindent
where  $\bar{\sigma} = - \sigma$  and $J \sim |\Gamma |^2 /
|^5E_0 -^4E_0-\mu |, \bar{J}\sim  |\Gamma |^2 / | ^5E_0 -^6E_0+\mu |$.

The dominant $p$ and $h$  processes are  diagonal in the
spin indices and involve only lead electrons with $\sigma $ {\sl opposite}
or {\sl equal }  to  $S_z$, respectively (see Fig.2).
The other possibility  corresponds to subleading  processes
in which  the intermediate dot state is a triplet.
$Y$ is the coupling strength for tunneling  processes that change both
$M$ and $S_z $ (off-diagonal terms in the spin labels).

\begin{figure}
\centering \includegraphics*[width=0.6\linewidth]{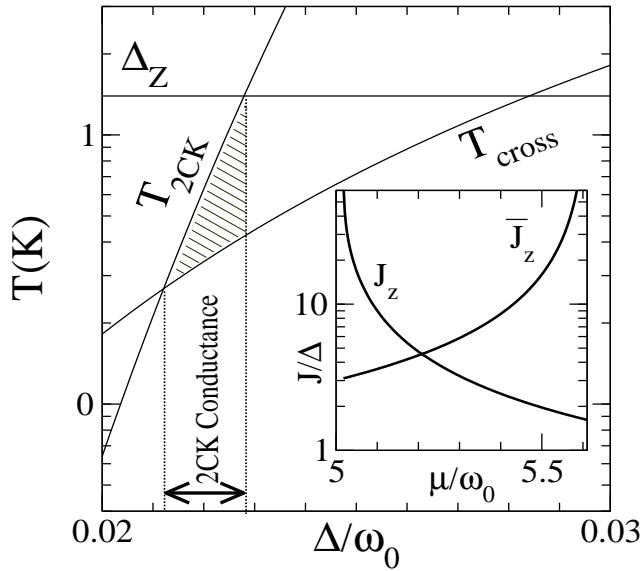}
\caption{ The plot shows $T_{\rm cross}$, $T_{2CK}$ and the Zeeman
spin splitting $\Delta_Z$ as a function of the hybridization parameter
$\Delta$, according to expressions given in the text.
The 2CK conductance may occur when $\Delta_Z > T_{2CK} > T_{\rm cross}$
(hatched area). Values of the parameters are: $\omega _0 = $6 meV,
$D_0$=30meV, $\Delta_Z =$ 1.4 K, $\nu(0)=$0.084meV$^{-1}$.  Inset:
 Plot of the couplings $J_z$  and $\bar{J}_z $ when the  position
of the dot levels relative to   $\mu$  is shifted by changing $V_g$.}
\label{tre}
\end{figure}

We calculate the couplings in $H_{\rm Eff}$ by means of exact numerical
diagonalization (Inset of Fig.(\ref{tre}) ). 
The transverse and longitudinal amplitudes, $(J^z,
J^{\perp}),(\bar{J}^z, \bar{J}^{\perp}), (Y^z,
Y^{\perp}$), are all of the same order of  magnitude. Anisotropies between
transverse and longitudinal couplings are irrelevant to our argument
anyway. Therefore, we assume isotropic magnetic couplings in
$H_{\rm Eff}$.

{\sl  $ T > T_{\rm cross} > \Delta _Z $ case :}

A scaling analysis on $H_{\rm Eff}$ is performed by rescaling the band
cutoff as $D \to D - \delta D$ and by working out the corresponding
renormalization of the coupling strengths $J, \bar{J}$ and $Y$. This
provides third order scaling equations given by:
\begin{eqnarray}
\frac{ d J}{ d \ln (\frac{T}{D_0})}  = \{
-  \nu ( 0 ) [ J^2 + Y^2 ] +  \nu^2 ( 0 ) [ J^3 + J Y^2 ] \} \;\; ;
\nonumber\\
\frac{ d \bar{J} }{ d \ln (\frac{T}{D_0})} = \{
-  \nu ( 0 ) [ \bar{J}^2 + Y^2 ] +  \nu^2 ( 0 )
[ \bar{J}^3 + \bar{J} Y^2 ] \}
\;\; ;
\nonumber\\
\frac{ d Y}{ d \ln (\frac{T}{D_0})} = \{ -  \nu ( 0 ) [ J Y + \bar{J} Y ]
+  \nu^2 ( 0 ) [ J^2 + \bar{J}^2 ] Y \}
\label{scaeq}
\end{eqnarray}
\noindent
($\nu ( 0 ) $ is the density of states at $\mu$ for each spin polarization).

The initial values of the parameters  $J_0$, $ \bar{J}_0$, $ Y_0$  define
a crossover temperature to strong coupled regime, $T_{\rm cross}$.
Because of the spin flip term proportional to $Y$ in Eq.(\ref{eff4}), 
diagonalization of $H_{\rm Eff}$ selects one preferred scattering channel,
which does not conserve spin and orbital momentum separately. Hence, below
$T_{\rm cross}$, the system flows towards a 1-channel underscreened 
strongly-coupled state. Given  isotropic  couplings,  
$J_0\sim \bar{J}_0 \sim Y_0$, $T_{\rm cross} = D_0 \sqrt{  \nu ( 0 ) J_0 }
\exp [ -\frac{1}{ 2 \nu ( 0 ) J_0 } ] $. 

{\sl $\Delta _Z \gg T_{\rm cross} $ case :}

Instead, we are interested in the 
complementary situation, $\Delta _Z > T > T_{\rm cross} $. The two 
$S_z = -\met $ dot's states decouple from the scaling when $ T < \Delta_Z$, and
spin-$\frac{1}{2}$ 2CK may take place in the system.
At temperatures below the decoupling point  $H_{\rm eff}$ no longer applies. 
Indeed, the off-diagonal term proportional to
$Y$ in Eq.(\ref{eff4}) does not contribute to scaling anymore. As the
dot lays within either state with $S_z = \frac{1}{2}$, $\vec{S}_{\uparrow
\uparrow}$ becomes a spin $\frac{1}{2}$-operator, $P_{\uparrow \uparrow} =
\frac{1}{2}$, while $P_{\downarrow \downarrow} = 0$. The effective Hamiltonian
describing low-energy dynamics at the scale $\Delta_Z$ takes the form
of a  2CK-Hamiltonian $H_{2CK}$,  with the spin index playing the role of
the channel index. Coupling constants in $H_{2CK}$ are given by
the coupling strengths $J, \bar{J}$ scaled down to $ T = \Delta_Z$,
$J^*, \bar{J}^*$. After a particle-hole transformation  is   performed
in the  $\uparrow$-spin channel  only  ($ b_{q m\uparrow}
\leftrightarrow  (-1)^{\bar{m}} b^\dagger_{q \bar{m} \uparrow}$, with
$m=0,2$  and $\bar{m} \neq m $), the Kondo interaction term in $H_{2CK}$ is
unchanged, while a sign is reversed in the potential part,  yielding:
\beq
H_{2CK} = \met (J^*  \rho_{\uparrow \uparrow} +  \bar{J}^*
\rho_{\downarrow \downarrow} )
+ J^* \vec{S} \cdot \vec{m}_{\uparrow \uparrow} + \bar{J}^*
\vec{S}\cdot \vec{m}_{\downarrow \downarrow}
\label{hef2}
\eneq
\noindent

As shown in the inset of Fig.3, fine-tuning of dot's parameters is possible,
to achieve
the symmetry condition $\bar{J}^*= J^*$.
Due to the transformation all the virtual processes are all of the same kind 
and the label $p, h$ has been dropped. The potential scattering 
in Eq.(\ref{hef2}) generates a phase shift responsible for depletion of 
$p$ states at the Fermi surface. For this reason, it has to be taken into 
account by diagonalizing it at each step of the renormalization
process described by the flow equation:

\beq
\frac{ d  J^*  }{ d \ln ( \frac{T}{\Delta_Z} )} =
- \nu^p(0) ( J^*)^2 + ( \nu^p(0) )^2  (J^* )^3 \;\;\; ,
\label{rg2c}
\eneq
\noindent
($\nu^p(0)$ is the $p$ density at the Fermi level) \cite{mfabri}.
From the scaling equation Eq.(\ref{rg2c}), one
derives the corresponding 2CK-temperature:
$T_{2CK} =  \Delta_Z \nu^p(0) J^* \: e^{- \frac{1}{ \nu^p(0) J^*}}$.
Below, we show that there is a wide range of parameters for which
the condition $T_{2CK} > T_{\rm cross}$ is realized. In this case, as
$T \to 0$, the system may flow toward a ``spin-1/2'' 2CK
fixed point.
Indeed, spin $\met $ 2CK  takes place if
coupling  of the two conduction channels is symmetric,
$\bar{J}^*= J^*$.   Although this condition  is  quite demanding,
   it should  be possible to  achieve it
by properly tuning  the gate voltage $V_g$, as
 shown in the inset of Fig.3.  Inclusion of the excited states we
neglected in our work ($N=4,6$ triplets and so on), will shift
the value of  $V_g$ corresponding to  the symmetry point.
 The signature of 2CK
spin $\met $ fixed point will be in a $\sqrt {T}$ dependence of the
conductance below
$T_{2CK}$  at a special $V_g $ point, instead of   the $T^2$   behavior,
 typical of a Fermi liquid
fixed point \cite{affleck}(see Fig.1 $b)$).
In Fig.3 we plot the corresponding
relevant temperature scales, $T_{2CK}$ and $T_{\rm cross}$ vs. the
 hybridization parameter between dot and leads, $\Delta = \nu ^p(0)
|\Gamma |^2$, in units of the single particle level spacing $\omega_0
= \sqrt{ \omega _d ^2 + \omega _c ^2/4 } = $6 meV  ($\omega_d = 4 meV$
is the frequency associated to the confining potential and
  $\omega _c = 9 meV$ is the cyclotron
frequency at $B=B^* $). Appropriate choice of the barrier thickness can fix
the hybridization parameter $\Delta$ within the required range
$\Delta _Z > T_{2CK} > T_{\rm cross} $.

In conclusion, we propose to search for orbital ``spin $\met $''  2CK in a
QD in a pillar  structure, tuned at CB with an odd number of
electrons  and an appropriate magnetic field along the axis, corresponding
to the degeneracy point between GS levels with  different values of $M$.
Detecting the effect requires an appropriate control of the hybridization 
of the dot with the contacts and a proper tuning of
the gate voltage  $V_g$. A measurement of the conductance $G(T)$ vs. $T$ 
at low temperature will exhibit quadratic behavior and will
cross  over to a square root behavior, as $V_g$ is tuned so
that the two channels become symmetric.

\acknowledgments
We gratefully acknowledge fruitful discussions with
I.Aleiner, B.Altshuler  and M. Fabrizio. Work partially supported by
TMR Project, contract FMRX-CT98-0180.


\begin{thebibliography}{99}
\bibitem{parity} We understand that we have performed a canonical
transformation from the usual basis set, where contact state operators
are labeled as e.g. Left (L) and Right (R), to a basis set given by
even- and odd-parity linear combinations with respect to the dot's position.
We also assume that the dot's potential is even, so that the two parities
decouple, and that the potential itself is strongly localized, so that
odd-parity combinations are ruled out.
\bibitem{nota}
Contact operators in Eq.(\ref{mag})  cannot be expressed in terms of
quadratic
combinations of operators $b_{qm\sigma}$ and $b^\dagger_{qm\sigma}$
since ordering in Eq.(\ref{mag}) is fixed.
\end{thebibliography}
\end{document}